\documentclass[a4paper,11pt]{article}
\usepackage{pos}
\newcommand{\beq}{\begin{equation}}
\newcommand{\eeq}{\end{equation}}

\title{Hadronization in small and large systems}

\author*[a]{Andrea Beraudo}

\affiliation[a]{Istituto Nazionale di Fisica Nucleare - Sezione di Torino,\\
  Via Pietro Giuria 1, 10125 Torino, Italy}

\emailAdd{beraudo@to.infn.it}

\abstract{Recent results on particle production in hadronic collisions at the LHC, from proton-proton (pp) to nucleus-nucleus (AA), challenge the traditional paradigm of hadronization as a universal late-time process which can be factorized from the partonic description of the rest of the event. If the violation of this description in nuclear collisions has been accepted for long -- with several hadronic observables finding a natural interpretation assuming some form of recombination of partons from a deconfined medium -- this same occurrence came as a surprise in proton-proton collisions. Actually, strong indications that hadronization cannot be simply described by universal fragmentation fractions/functions were already found in fixed-target experiments, but limited to specific kinematic regions close to the beam rapidity, so that a complete change of paradigm did not look necessary. Here we show that a wide set of observables finds a natural interpretation assuming that in all high-energy hadronic collisions, from pp to AA, a deconfined medium is formed, acting as a reservoir in which a parton can undergo local color neutralization (LCN) through recombination with an opposite color charge. Hadronization, belonging to the non-perturbative domain of QCD, will always, inevitably, require some amount of educated modelling. However, some commong features to the various proposed model in the literature can be identified, which look necessary in order to describe the most recent data. Here we decide to focus on heavy-flavor (HF) production, since in this case the origin of at least one of the consituent quarks of the final hadron is well known and attributed to an initial hard partonic process.}

\FullConference{12th Large Hadron Collider Physics Conference  (LHCP2024)\\
3-7 June 2024 \\
Boston, USA \\}

\begin{document}
\maketitle


\begin{figure*}
\centering
\includegraphics[clip,height=5cm]{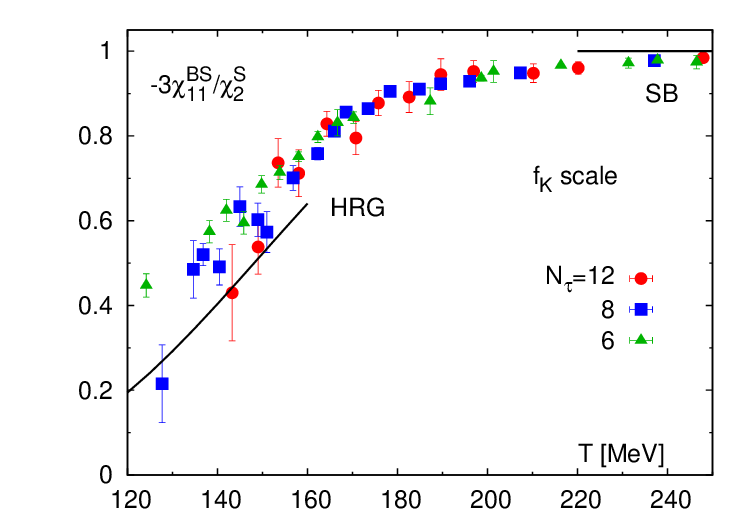}
\includegraphics[clip,height=5cm]{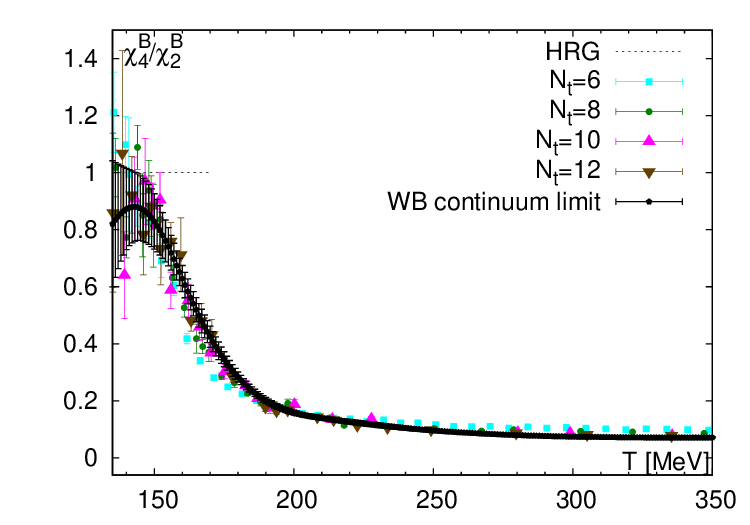}  
\caption{Ratios of different generalized susceptibilities of conserved charges in $N_f=2+1$ lattice-QCD simulations~\cite{HotQCD:2012fhj,Borsanyi:2013hza}. The parton-to-hadron transition in the nature of charge carriers looks very smooth.}\label{fig:lQCD}
\end{figure*}
Providing predictions for particle production in high-energy collisions requires facing the challenge of modelling hadronization, going from a partonic description of the event to the final hadronic degrees of freedom actually measured in the detectors. For the sake of simplicity this transition is usually described as taking place suddenly: for instance, in a QCD shower this occurs once a parton emerges from a branching with a virtuality below a minimum value $Q_0\sim 1$ GeV; if, on the other hand, hadrons are assumed to decouple from a hot fireball particle abundances turns out to be described by a chemical freeze-out temperature $T_H\approx 160$ MeV, compatible with the QCD crossover temperature found by lattice simulations. Any discussion concerning hadronization must start from the caveat that its description as a sudden instantaneous process is an oversimplification. What occurs during hadronization is a change in the nature of the carriers of conserved charges, specifically -- in the $N_f=2+1$ case -- baryon number ($B$), electric charge ($Q$) and strangeness ($S$). As one can see in Fig.~\ref{fig:lQCD}, in which different ratios of generalized susceptibilities measured on the lattice are shown, this change is actually very smooth. For instance, in the deconfined phase each particle carrying one unity of strangeness has $B=-1/3$, while in a hadron gas strangeness is mostly carried by kaons, with $B=0$: accordingly one would expect a sharp transition from 0 to 1 in the left panel of Fig.~\ref{fig:lQCD} which is not observed. The same is true for the ratio of the forth to second-order cumulants of baryon number fluctuations, which should be proportional to $B^2$: no discontinuous jump from 1 to 1/9 is observed, consistently with the smooth nature of the QCD (de)confinement crossover.

\begin{figure*}
\centering
\includegraphics[clip,height=5cm]{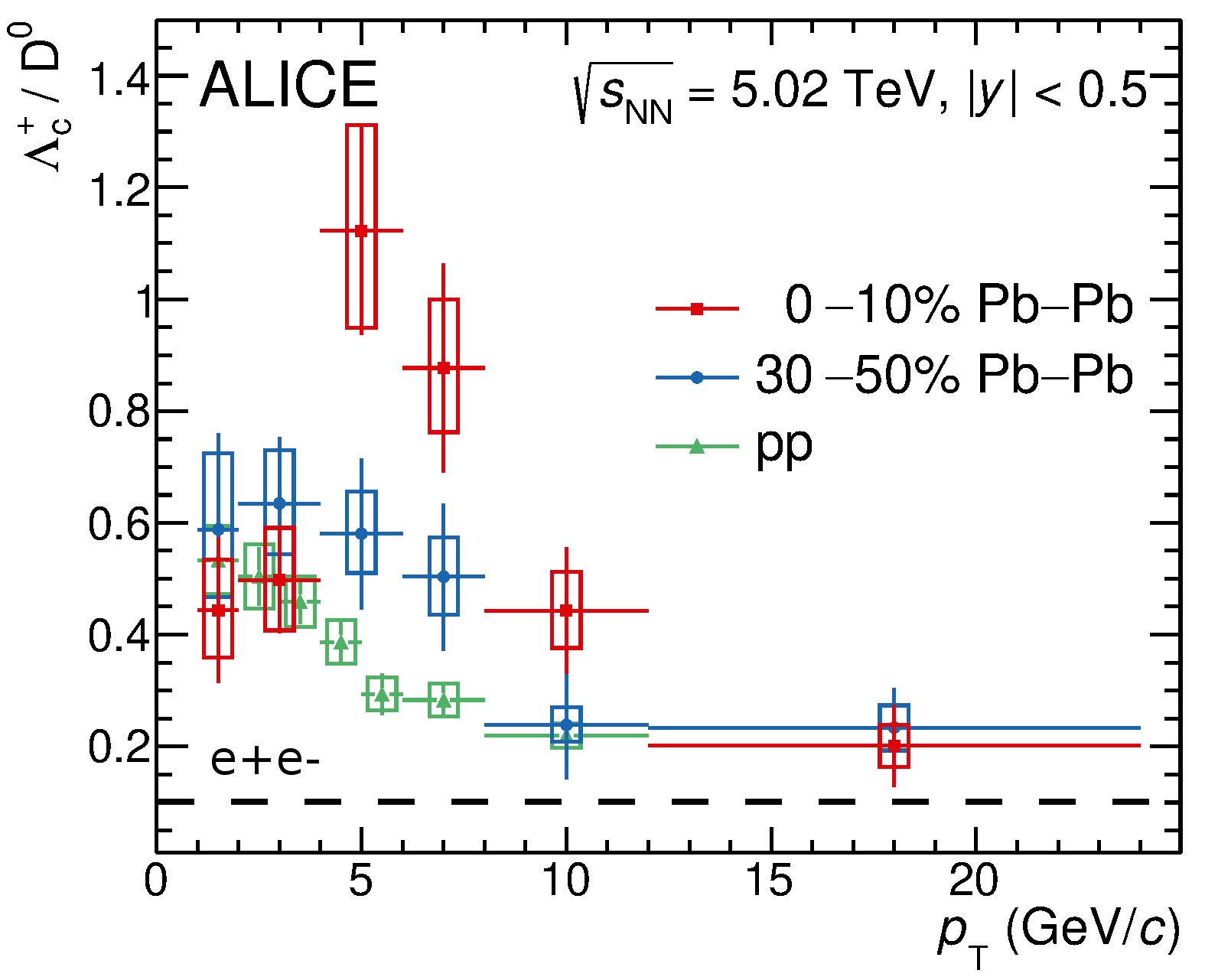}
\includegraphics[clip,height=5cm]{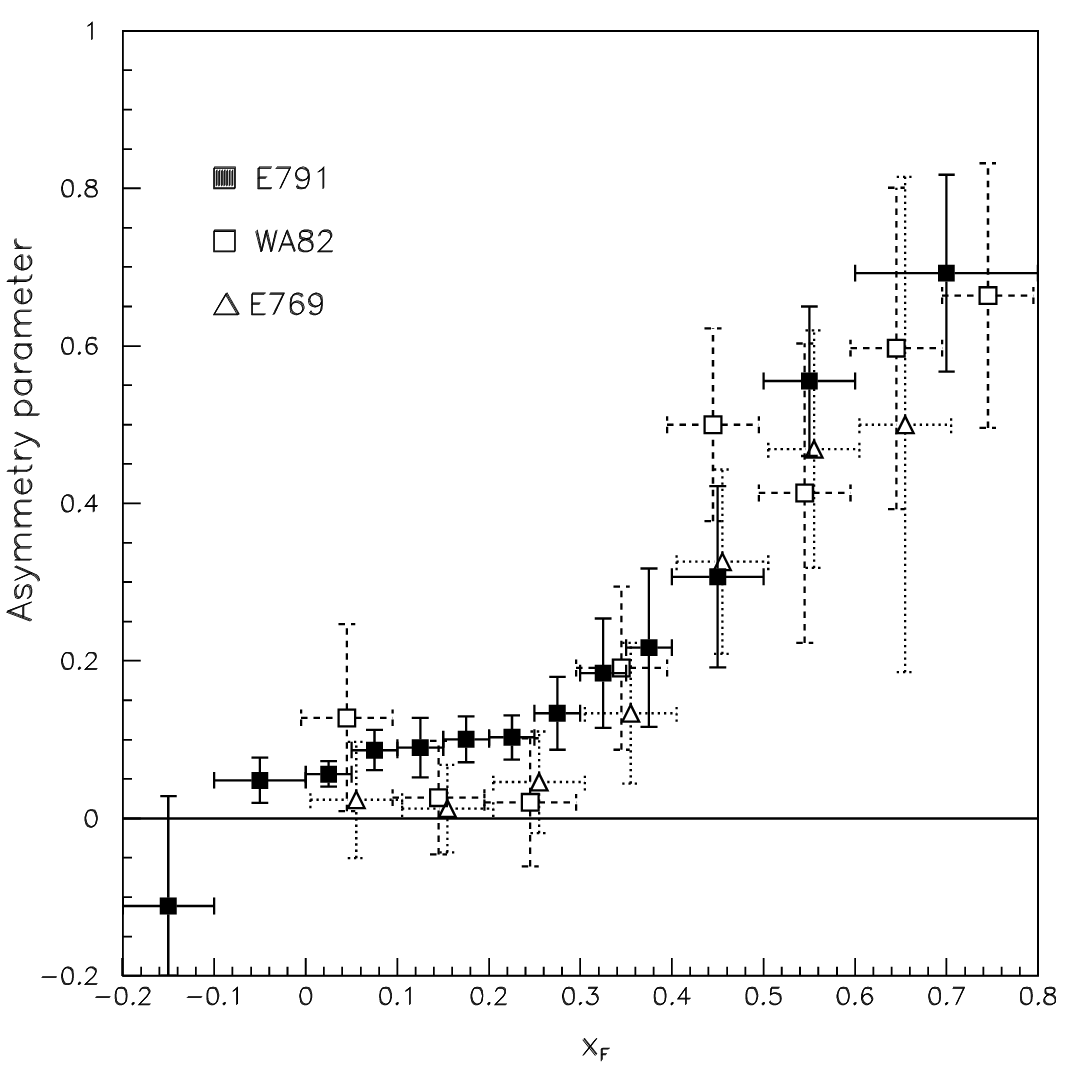}  
\caption{Left panel: $\Lambda_c^+/D^0$ ratio from minimum-bias pp to central Pb-Pb collisions at $\sqrt{s_{\rm NN}}=5.02$ TeV~\cite{ALICE:2020wfu}. Right panel: asymmetry in the $D^+$ vs $D^-$ meson production in $\pi^-$-nucleon collisions at various fixed-target experiments~\cite{E791:1996htn}.}\label{fig:exp-data}
\end{figure*}
Hadron production in a hard QCD collision is traditionally described within the framework of the so-called factorization theorem, according to which a hadronic cross-section can be obtained from the convolution of a partonic one with the parton distribution functions (PDF's) of the two projectiles and the fragmentation function (FF) into to the final given hadron. If factorization holds, PDF's and FF's are non-perturbative universal objects: they cannot be calculated from first principles, but once measured in a process ($e^+e^-$ in the case of FF's) they can be applied also to other collisions. However, several recent results concerning heavy-flavor hadron production challenge the above interpretation. For instance, a strong enhancement by a factor 5 of the $\Lambda_c^+/D^0$ ratio -- both for the $p_T$-spectra and for the integrated yields -- was observed in pp collisions with respect to expectations based on $e^+e^-$ results~\cite{ALICE:2020wfu}. Actually asymmetries in charmed-particle production in hadronic collisions were already observed in fixed-target experiments at SPS and Fermilab (see for instance~\cite{E791:1996htn}), but limited to quite differential measurements in a kinematic regime close to the beam rapidity. The fact that now the integrated yields themselves deviate from the expectation based on $e^+e^-$ collisions entails a violation of the factorization theorem,
\beq
d\sigma_h{\ne}\sum_{a,b,X}f_a(x_1)\,f_b(x_2)\,\otimes d\hat\sigma_{ab\to c\bar cX}\,{\otimes D_{c\to h_c}(z)},
\eeq
since the FF's $D_{c\to h_c}(z)$ cannot be considered universal any longer, but depend on the environment in which hadronization takes place.

\begin{figure*}
\centering
\includegraphics[clip,height=5cm]{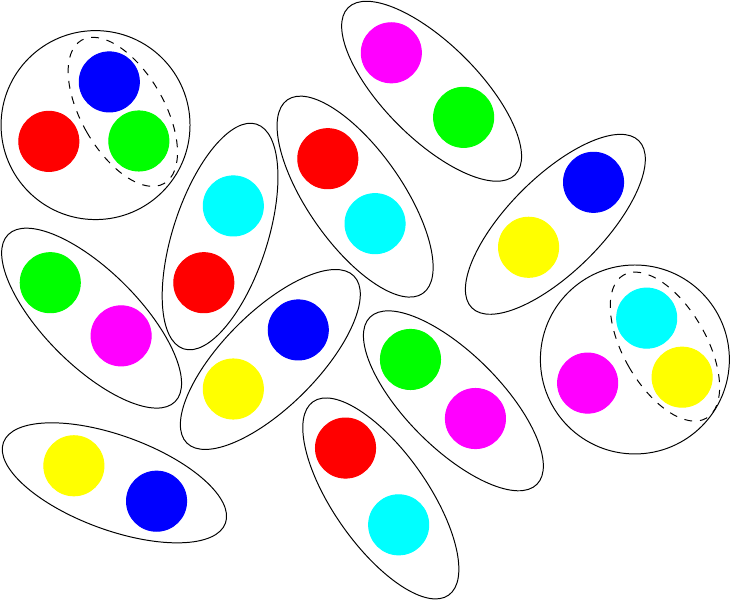}
\includegraphics[clip,height=5cm]{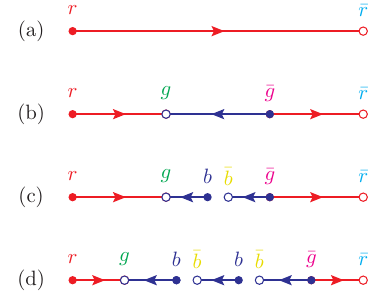}  
\caption{Mechanism of baryon production withing the Local Color Neutralization mechanism (left panel) described in~\cite{Beraudo:2022dpz} compared to the Pythia string-breaking popcorn model implemented in Pythia~\cite{Bierlich:2023okq}.}\label{fig:LCNvsPYTHIA}
\end{figure*}
The fact that the enhanced charmed baryon production (not limited to $\Lambda_c^+$ baryons) observed in pp collisions looks comparable to the one in the Pb-Pb case suggests that a common hadronization mechanism is at work in both situations. One can imagine, as illustrated in the left panel of Fig.~\ref{fig:LCNvsPYTHIA}, that in any high-energy hadronic collision a small deconfined fireball is formed, acting as a reservoir from which a (charm, in this case) quark can find an opposite color charge to recombine with, leading to the formation of a color-singlet cluster (see Ref.~\cite{Beraudo:2022dpz} for details). It is reasonable to assume that recombination takes place locally, involving the closest opposite color charge (Local Color Neutralization, LCN in the following), which can be either an antiquark or a diquark in case the quark-quark interaction in the antitriplet channel is sufficiently attractive: this last occurrence favors the formation of clusters carrying $|B|\!=\!1$, naturally explaining the enhanced baryon-to-meson ratio observed in hadronic collisions.
The fact that recombination occurs locally within a fireball undergoing a collective expansion entails that the two particles, beside being close in space, have also quite collinear momenta. This gives rise to the formation of low invariant-mass clusters, typically decaying into just two final-state particles which, furthermore, inherit the collective flow of the fireball. Notice that in $2\to 1$ coalescence approaches (see Ref.~\cite{Fries:2008hs} for a review) the singlet cluster arising from the recombination is directly identified with a final-state hadron/resonance. It is interesting to compare the above picture with the hadronization mechanism implemented in the Pythia event generator, based on the string-fragmentation model, in particular concerning the issue of baryon production (see Ref.~\cite{Bierlich:2023okq}). Notice, first of all, that in Pythia color neutralization is not local, but usually occurs through the formation of quite elongated strings of large invariant mass, connecting partons quite far in rapidity. String breaking occurs through the excitation of $q\bar q$ pairs from the vacuum within the strong chromoelectric field between the two endpoints. The excitation of multiple $q\bar q$ pairs can give rise to pairs of nearby strings with (anti)quark-(anti)diquark endpoints (popcorn mechanism). This leads to the production of baryon-antibaryon pairs very close in rapidity or at most separated by one meson: baryon number (and other charges as well) conservation occurs \emph{locally} in standard Pythia implementations. On the contrary, in the LCN model such a constraint is absent: the formation of a baryon in a given fluid cell can be compensated by the production of an antibaryon from the opposite side of the fireball. Experimental measurements of net baryon-number fluctuations~\cite{ALICE:2019nbs} allows one to determine the volume over which the latter is conserved, favoring a scenario with long range $B-\overline B$ correlations, in contradiction with a simple string-fragmentation model.

\begin{figure*}
\centering
\includegraphics[clip,height=4cm]{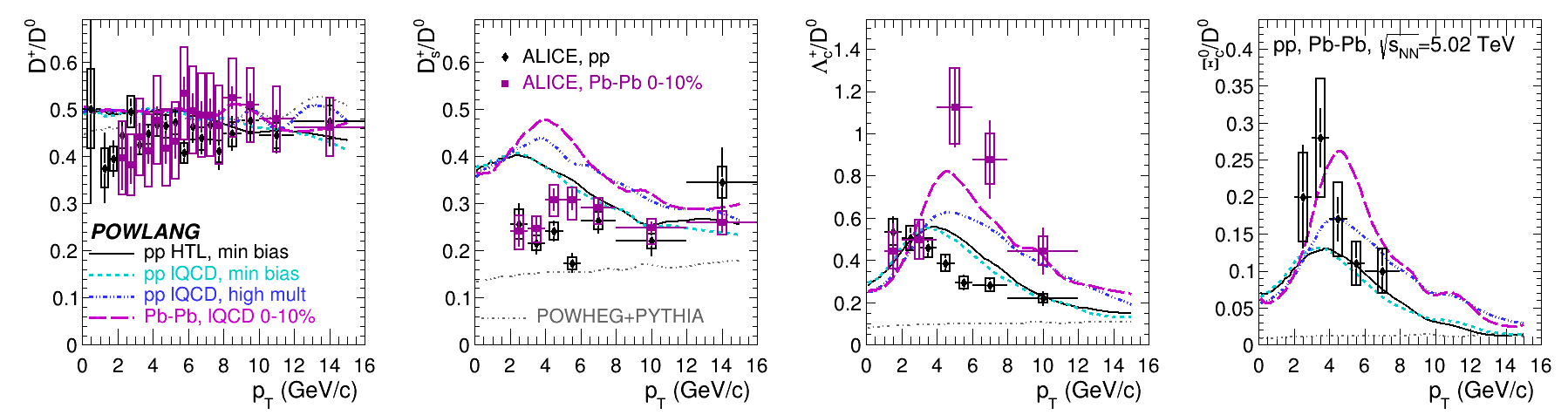}  
\caption{Charmed-hadron yield ratios as a function of $p_T$ for different hadronic collisions, from minimum-bias pp to central Pb-Pb, at $\sqrt{s_{\rm NN}}\!=\!5.02$ TeV. Predictions including in-medium transport supplemented with the LCN hadronization model~\cite{Beraudo:2023nlq} are compared to ALICE data~\cite{ALICE:2020wfu,ALICE:2021mgk,ALICE:2021rxa,ALICE:2021bib,ALICE:2021psx}.
Also shown are the pp predictions obtained with POWHEG+PYTHIA standalone~\cite{Alioli:2010xd}, with no medium effects, which undershoot charmed-baryon production.}\label{fig:LCNresults}
\end{figure*}
It is interesting to study the predictions of the LCN model for heavy-flavor hadron production across different colliding systems, from minimum-bias pp to central Pb-Pb. In Fig.~\ref{fig:LCNresults} the results for the $p_T$-dependent charmed-hadron yield ratios are shown~\cite{Beraudo:2023nlq}. Theoretical curves were obtained assuming that in all systems a hot deconfined fireball is formed, affecting both the propagation of charm quarks (described by transport equations) and their eventual hadronization. Differences among the various hadronic collisions are due to the size and lifetime of the fireball, which can affect the collective flow inherited from the medium, without however leading to substantial differences in the hadrochemistry: in all cases the integrated yields display an enhanced production of charmed baryons with respect to the hypothesis of universal FF's extracted from $e^+e^-$ collisions.

\begin{figure*}
\centering
\includegraphics[clip,height=5cm]{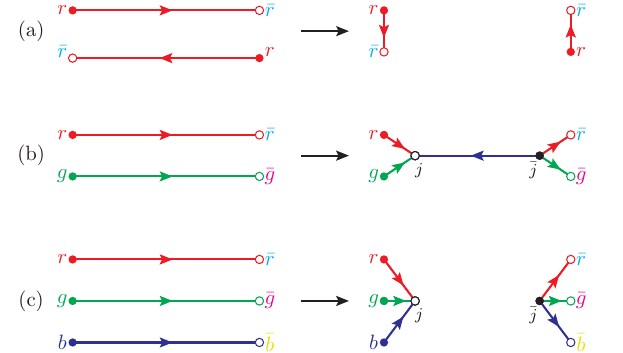}
\includegraphics[clip,height=5cm]{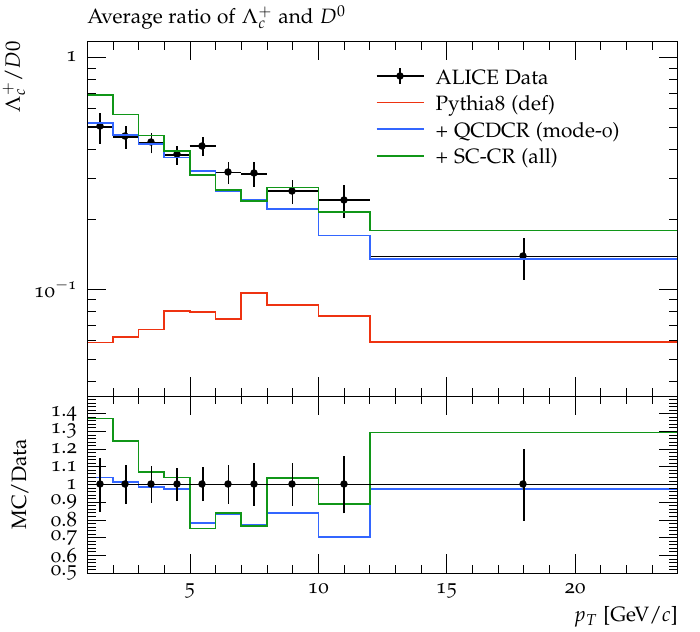}  
\caption{Cartoon of the CR model implemented in PYTHIA (left panel) leading to an enhanced baryon production~\cite{Bierlich:2023okq}, in good agreement with the experimental data for the $\Lambda_c^+/D^0$ ratio (right panel).}\label{fig:CR}
\end{figure*}
How can a more realistic modelling of hadronization be included into multi-purpose QCD event generators? A promising strategy is the implementation of the so-called Color Reconnection (CR) mechanism, currently included in the most recent versions of Pythia and Herwig~\cite{Gieseke:2017clv}. A schematic picture is displayed in the left panel of Fig.~\ref{fig:CR}. A rearrangement of color connections is allowed if this leads to the formation of smaller invariant-mass color-singlet structures and new topologies (baryon junctions) are possible. In particular, this leads to an enhanced probability of baryon production, no longer requiring the excitation of diquark-antidiquark pairs from the vacuum, allowing one to get a more satisfactory agreement with the experimental data~\cite{Bierlich:2023okq}, as shown in the right panel of Fig.~\ref{fig:CR}. As one can see in Fig.~\ref{fig:CR} CR makes color neutralization more local, the invariant mass of the new color-singlet objects is smaller and baryon-number conservation does not need to occur locally (see panel c). All the above features makes these updated hadronization routines quite similar to the LCN mechanism discussed in the previous paragraphs and look necessary ingredients for any hadronization model aiming at describing the experimental data.

\bibliographystyle{JHEP}
\bibliography{beraudo-LHCP}


\end{document}